\documentclass[aps, pra, twocolumn, amsmath]{revtex4}

\usepackage{graphicx}
\usepackage{subfigure}
\usepackage{amssymb}
\usepackage{bm}
\usepackage{here}

\newcommand{\vect}[1]{\bm{#1}}

\newcommand{\Bra}[1]{\left\langle{#1}\right\rvert}
\newcommand{\Ket}[1]{\left\lvert{#1}\right\rangle}

\newcommand{\mgh}{MgH$^{+}\,$}
\newcommand{\arh}{ArH$^{+}\,$}

\newcommand{\oh}{OH$^+\,$}

\newcommand{\bh}{BH$^+\,$}
\newcommand{\sh}{SH$^{+}\,$}

\newcommand{\nh}{NH$^{+}\,$}
\newcommand{\hf}{FH$^{+}\,$}
\newcommand{\triplet}{$^3\Sigma\;$}
\newcommand{\doublet}{$^2\Sigma\;$}
\newcommand{\singlet}{$^1\Sigma\;$}
\newcommand{\doubletpi}{$^2\Pi\;$}

\newcommand{\Braket}[1]{\left\langle{#1}\right\rangle}

\begin{document}
\title{Rotational cooling of heteronuclear molecular ions with \singlet, \doublet, \triplet and \doubletpi electronic ground states}
\author{I. S. Vogelius}
\affiliation{Department of Physics and Astronomy,
  University of  Aarhus, 8000 {\AA}rhus C, Denmark}
\author{L. B. Madsen}
\affiliation{Department of Physics and Astronomy,
  University of  Aarhus, 8000 {\AA}rhus C, Denmark}
\author{M. Drewsen}
\affiliation{Danish National Research Foundation Center for
  Quantum Optics and Department of Physics and Astronomy, University of Aarhus, 8000 {\AA}rhus C, Denmark}

\date{\today}

\begin{abstract}
 The translational motion of molecular ions can be effectively cooled sympathetically to translational temperatures below 100 mK in ion traps
 through Coulomb interactions with laser-cooled atomic ions. The ro-vibrational degrees of freedom, however, are expected to be largely
 unaffected during translational cooling. We have previously proposed schemes for cooling of the internal degrees of freedom of such
 translationally cold but internally hot heteronuclear diatomic ions in the simplest case of \singlet electronic ground state molecules. Here we present a significant simplification of these schemes and
 make a generalization to the most frequently encountered electronic ground states of heteronuclear molecular ions: \singlet, \doublet, \triplet
 and \doubletpi. The schemes are relying on one or two laser driven transitions with the possible inclusion of a tailored incoherent far
 infrared radiation field.

\end{abstract}

\maketitle


\section{Introduction}
The cooling and manipulation of neutral molecules has become the subject of intense studies in recent years and impressive advances have been
made. Experiments include the successful production of molecular Bose-Einstein condensates
\cite{GrimmMolecularBEC,KetterleMolecularBEC,JinMolecularBEC}, the deceleration and trapping of polar molecules in inhomogeneous fields
\cite{meijer3,meijer1,meijer2,MeijerReview} and loading a trap with paramagnetic molecules cooled by a He buffer gas \cite{weinstein,egorov}.
For the NH radical the presence of an unusually large Franck-Condon factor offers prospects for direct Doppler cooling of a trapped molecule
\cite{meijer2003}.

Molecular ions constitute another class of molecules that are very interesting to cool and manipulate. Diatomic molecular ions are, e.g.,
important constituents of interstellar media \cite{IAUsymposiumProceedStellarMolecIons,IAUsymposiumProceedCometsMolecIons}, comets and cool
stellar atmospheres including that of the sun \cite{IAUsymposiumProceedCometsMolecIons,ComettailOh}, and there have recently been proposals to
utilize cold molecules to implement a quantum computer \cite{demille,QuantCompVibrationalLevelsMolec_Tesch}.

The cooling of molecules is in general more complicated than that of atoms since the ro-vibrational substructure of the electronic molecular
energy levels normally makes it impossible to find a closed optical pumping scheme to be used for conventional laser cooling. Molecular ions,
however, may be very effectively cooled sympathetically by loading them into a trap with laser cooled atomic ions
\cite{molhave,DrewsenIntJounMassSpectrArticle2003,SchillerDusseldorfMolecularDynamicsSimulationInPaulTrap,SchillerBeTrappingPreliminary,SympatheticCoolingMoleculesInPenning}.
The Coulomb interaction between the charged particles provides efficient momentum transfer from the initially hot molecular ions to the cooled
atomic ions. Dissipative cooling of the translational motion is hence obtained for both species although only the atomic ions are subject to
laser cooling.

One might expect that the ro-vibrational degrees of freedom of a diatomic molecule placed in the vicinity of a cooled atomic ion would couple to
the translational motion of the atomic ion, resulting in strong sympathetic cooling of these degrees of freedom. In a typical ion trap, however,
the excitation energy of the translational atomic motion in the trap (vibrations in the harmonic trap potential) is of the order of 1 MHz, which
is much smaller than typical energies of ro-vibrational excitations (of the order $10^{11}-10^{14}$ Hz). The large difference between these
numbers prohibits that the internal ro-vibrational states couple effectively to the external motion of the ions in the trap. In the following we
therefore assume that the internal degrees of freedom relax to equilibrium with the black-body radiation (BBR) present in the trap. This will
happen on a timescale of tens of seconds, which is significantly faster than the inelastic collision time in the trap which, from Langevin
theory, is estimated to be hundreds of seconds \cite{vogelius}.

In Refs.~\cite{vogelius,vogeliusOptimizeLamp_JPhysB} we proposed schemes for cooling of the rotational degree of freedom of such molecular ions
in the case of heteronuclear molecules with a \singlet electronic ground state. The schemes are based on two direct infrared (IR) transitions
between the lowest vibrational states in the molecule or two Raman transitions coupling the vibrational levels via a near-resonant excited
electronic state. In addition to the pumping by the external light sources the cooling schemes are assisted by rotational redistribution
mediated by the BBR. The timescale of the cooling schemes are on the order of $\sim60$ s which is shorter than the estimated inelastic collision
rate with background gas.

Though most molecules appearing in nature have a \singlet electronic ground state it is necessary to consider other electronic states for
molecules produced in the laboratory, including molecular ions. The by far most frequently encountered electronic ground states of such
molecules and molecular ions are, apart from the \singlet state, the \doubletpi, \doublet and \triplet states. This includes the lighter
diatomic hydrides, e.g., \hf (\doubletpi), \bh (\doublet) and \oh (\triplet). Such ionized hydrides are attractive candidates for our cooling
schemes as they have low reduced masses and hence high rotational transition frequencies leading to fast rotational relaxation rates which is
beneficial for the timescale of the cooling scheme. By extending the schemes to \doubletpi, \doublet and \triplet states we have then covered
all the lighter ionic hydrides and the vast majority of other molecules amenable for cooling. We show that it is possible to cool such
electronic states, though at the cost of introducing more laser frequencies in some cases. For most molecules, however, the present cooling
schemes rely on only a single IR laser, possibly assisted by broadband radiation from a far-infrared (FIR) emitter which is filtered to optimize
the cooling efficiency.

The present paper is organized as follows. In Sec.~\ref{Sec:CoolingSchemeSinglet} we present cooling schemes for the \singlet electronic ground
states. In Sec.~\ref{sec:NumericalSimSinglet} we discuss a model of the cooling schemes and present numerical simulations for \mgh
(X$^1\Sigma$). In Sec.~\ref{sec:rotationalsubstrcture} we present cooling schemes applicable to the \doublet, \triplet and \doubletpi electronic
ground states together with numerical simulations of each of the cooling schemes. A summary of the results is given in Sec.~\ref{sec:summary}.
In Appendix \ref{sec:AppendixEinsteinCoeffs}, we have collected the Einstein coefficients for the considered molecules and transitions, and in
Appendix \ref{sec:AppHonlLondon}, we describe the H\"{o}nl-London factors of interest.

\section{Cooling schemes for \singlet states} \label{Sec:CoolingSchemeSinglet}

\begin{figure}[H]
  \begin{center}
    \mbox{
      \subfigure[Cooling scheme of Ref.\ \cite{vogelius}]{\includegraphics[width=0.26\textwidth]{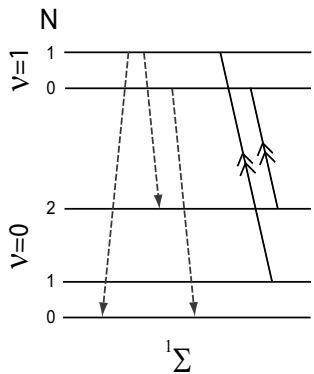}} }
    \\[0.1cm]
    \mbox{
    \subfigure[Raman scheme]{\includegraphics[width=0.26\textwidth]{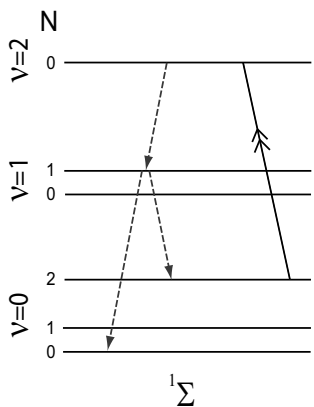}}
    }
    \\[0.1cm]
     \mbox{
          \subfigure[Direct scheme]{\includegraphics[width=0.26\textwidth]{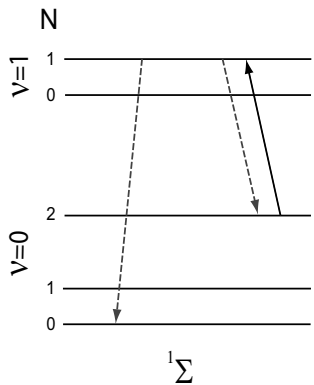}}
     }

  \caption{\label{fig:schemes} Ro-vibrational states of interest in the cooling schemes for \singlet states. The cooling concept involves
transitions between ro-vibrational states driven by Raman pulses (solid lines, double arrows in (a) and (b)) via an excited electronic state or
direct laser pumping (solid line, single arrow in (c)) and subsequent spontaneous decays (dashed lines).}
\end{center}
\end{figure}

The suggested schemes for \singlet states are sketched in Fig.\ \ref{fig:schemes}. The driven transitions are either Raman transitions via
  an excited electronic state or transitions directly between vibrational levels. Fig.~\ref{fig:schemes}(a) represents the cooling scheme of
  Ref.~\cite{vogelius} in which two Raman transitions make a closed cycle through pumping of population from the ``pump states'', $(\nu=0,N=1)$
  and $(\nu=0,N=2)$, to the excited states, $(\nu=1,N=1)$ and $(\nu=1,N=0)$, respectively, followed by subsequent spontaneous emission bringing
  the populations back to the ``pump states'' or to the ro-vibrational ground state. Here $\nu$ and $N$ denotes the vibrational and rotational
  level respectively. Population initially in higher-lying states is fed to the pump states through BBR--induced rotational transitions within
  the vibrational ground state.

  It would be advantageous for practical implementation to use only a single Raman (Fig.~\ref{fig:schemes}(b)) or a single direct
  (Fig.~\ref{fig:schemes}(c)) transition, at the expense of not emptying the $(\nu=0,N=1)$ state. Without applying other means to limit the
  pile-up of population in the $(\nu=0,N=1)$ state the cooling efficiency, measured as the percentage of population in the ground state, will
  decrease. One can, however, take advantage of the higher frequency of the $(\nu=0,N=1) \leftrightarrow (\nu=0,N=2)$ rotational transition
  compared to the undesired $(\nu=0,N=0)\rightarrow (\nu=0,N=1)$ heating transition inevitably driven by the BBR, and apply an incoherent source
  and a high-frequency pass filter to reduce the radiation resonant with the heating transition while still addressing the
  $(\nu=0,N=1) \leftrightarrow (\nu=0,N=2)$ transition. Thereby one can obtain the desired depletion of the $(\nu=0,N=1)$-population by means of
  incoherent radiation only. As the rate of depletion using realistic incoherent sources will be slower than if the state was addressed by a
  laser, it is necessary to design the cooling scheme such that spontaneous decays to the $(\nu=0,N=1)$ state from states which are
  participating in the pumping cycle are avoided. This can be done by addressing the $(\nu=0,N=2) \leftrightarrow (\nu=2,N=0)$ transition with a
  resonant, dipole allowed ($\Delta N=0,\pm 2$) Raman pulse as depicted in Fig.~\ref{fig:schemes}(b). The pumping to the $(\nu=2,N=0)$ state is
  then followed by spontaneous decays through $(\nu=1,N=1)$ to $(\nu=0,N=0)$ and $(\nu=0,N=2)$ in accordance with the dipole selection rules
  $((\Delta N,\Delta\nu)=\pm 1)$ for single photon decays.

  It is shown in Ref.~\cite{vogelius} that the Raman-transitions in the \mgh test case \cite{molhave} are saturated by a $\sim 100$ kW/cm$^2$,
  $10$ ns pulse, which is a modest intensity for present day laser systems.

One of our schemes using only a single direct laser-induced transition subject to the dipole selection rule is shown in Fig.\
\ref{fig:schemes}(c) \cite{vogeliusOptimizeLamp_JPhysB}. The laser pumps the $(\nu=0,N=2) \leftrightarrow (\nu=1,N=1)$ transition while
subsequent spontaneous decay brings the population back to the pump state or to the ro-vibrational ground state. A filtered incoherent source is
applied in order to bring population from the $(\nu=0,N=1)$ state to the "pump state". The advantage of this direct scheme is that it does not
depend on the existence of an excited electronic state that can be addressed with laser light and also that it requires only a single laser
frequency. \arh is an example of a molecule without excited electronic states \cite{ArHNoExcitedStateSchutte,vogelius}.

From a practical point of view, a pulsed laser system is desirable for the direct scheme of Fig.~\ref{fig:schemes}(c). The IR light could, for
example, be generated by difference frequency mixing of the primary beam of a frequency doubled Nd:YAG laser and a dye laser pumped with the
same beam. In the \mgh case, the wavelength of the $(\nu=0,N=2)\leftrightarrow (\nu=1,N=1)$ pumping transition is $\simeq 5.9 \ \mu$m
\cite{herzberg} and the Einstein A-coefficient is $\simeq 20$ s$^{-1}$. To ensure saturation of the laser driven transition we require that the
population in the states involved undergoes at least 10 Bloch oscillation during a laser pulse and that the amplitude of each oscillation
exceeds $0.9$. If we assume a detuning of $1$ GHz and a pulse duration $10$ ns we find that an intensity of $\sim 500$ W/mm${}^2$ is needed to
fulfill both requirements. This corresponds to a pulse energy of 5 $ \mu$J. Typical nonlinear crystals should be able to deliver an energy of
$\sim 10$ $\mu$J per laser pulse at the wavelength required.

The added incoherent field from a lamp will increase the rate of rotational transitions needed for cooling, but at the expense of heating the
population distribution. The spectral distribution of the incoherent field can therefore be shaped to maximize the cooling efficiency as
described in Ref.~\cite{vogeliusOptimizeLamp_JPhysB}.

\section{Numerical simulations for \singlet states} \label{sec:NumericalSimSinglet}
In this section, we present our model of the cooling scheme, show the results of numerical simulations and discuss the optimal radiation
distribution of the incoherent field.
\subsection{Rate equations for the population dynamics} \label{subsec:Rateeq}
The population dynamics is well-described by rate equations giving the change in population of a given state via Einstein coefficients and
frequency-specific radiation intensities. The equation of motion for the molecular population $P_i$ in state $i$ takes the form
\begin{equation}
\begin{split} \label{eq:rate}
\frac{dP_i}{dt}=&-\sum_{j=0}^{i-1}A_{ij}P_i + \sum_{j=i+1}^{M}A_{ji}P_j- \\
 &\sum_{j=0}^{i-1}P_i B_{ij}W(\omega_{ij})+\sum_{j=0}^{i-1}P_jB_{ji}W(\omega_{ij})-\\
 &\sum_{j=i+1}^{M}P_iB_{ij}W(\omega_{ij})+\sum_{j=i+1}^M P_j B_{ji} W(\omega_{ij}).
\end{split}
\end{equation}
Here
\begin{equation} \label{eq:defPopVector}
\begin{gathered}
\vect{P}=(P_{\nu=0,N=0},P_{\nu=0,N=2} \ldots P_{\nu=0,N=N_{max}}, P_{\nu=1,N=0}\\ \ldots P_{\nu=1,N=N_{Max}},P_{\nu=2,N=0}\ldots
P_{\nu=2,N=N_{max}})
\end{gathered}
\end{equation}
represents the populations in vector form with $N_{max}$ chosen so the population in higher-lying rotational states is negligible during the
cooling process. $A_{ij}$ and $B_{ij}$ are the Einstein coefficients describing spontaneous and stimulated transitions from energy level $i$ to
$j$. $W(\omega_{ij})$ is the cycle averaged radiative energy density present in the trap at the resonant transition frequency
$\omega=\omega_{ij}$, between level $i$ and $j$. In Eq.~\eqref{eq:rate}, the first term corresponds to spontaneous decay from state $i$ to
states with lower energy, while the second term describes spontaneous decay from levels with higher energy into state $i$. Stimulated emission
from the $i$th state and stimulated absorption from lower-lying states is then described by the third and fourth term, and finally, the last two
terms represent transitions due to absorption of radiation from the $i$th state and stimulated emission from higher-lying states into the $i$th
state.

The system of Eqs.~\eqref{eq:rate} is conveniently expressed by the matrix equation
\begin{equation} \label{eq:RateM}
\frac{d\vect{P}}{dt}=\bm K \vect{P},
\end{equation}
where $\vect{K}$ is an $(M+1) \times (M+1) $ coupling matrix.

\subsection{Calculation of molecular properties} \label{sec:SingletMolecularProperties}
As seen from Eq.~\eqref{eq:rate}, it is necessary to know the Einstein coefficients to simulate the population dynamics. For many molecules, the
Einstein coefficients are available in the literature. If not, they are evaluated numerically as follows. We use the well-known quantum
mechanical expressions for the Einstein coefficients between an upper state $\Psi_n$ and a lower state $\Psi_m$ that are both non-degenerate
\cite{loudon}

\begin{equation} \label{eq:QMEinsteinNonDegenerate}
\begin{split}
  B_{n,m} &= \frac{\pi |\vect{D}_{n,m}|^2}{3\epsilon_0 \hbar^2}, \\
  A_{n,m} &= \frac{\hbar \omega^3}{\pi^2 c^3}B_{n,m},
\end{split}
\end{equation}
where $\omega$ denotes the transition frequency and $\vect{D}$ the transition dipole moment between the states.
\begin{equation} \label{eq:Defdipolemoment}
    \vect{D}_{n,m}^{lab}=\int \Psi_n^{\ast} \vect{M}^{lab} \Psi_m d \tau,
\end{equation}
with $d\tau$ denoting the volume element corresponding to integration over the complete set of coordinates for all particles involved and
\begin{equation} \label{eq:defM}
\vect{M}^{lab}=\sum_{k} -e \vect{r}_k + \sum_{l=1,2} Z_l e \vect{R}_l
\end{equation}
the dipole operator.

The equations refer to a laboratory-fixed coordinate system so the molecular wave-functions include the rotational terms. The summation indices
in Eq.~\eqref{eq:defM}, \textit{k} and \textit{l}, refer to the electrons and the involved nuclei, while $Z_l$ denotes the nuclear charge.

For degenerate states Eq.~\eqref{eq:QMEinsteinNonDegenerate} is modified to
\begin{equation} \label{eq:QMEinstein}
\begin{split}
  B_{n,m} &= \frac{\pi |\mathcal{D}_{n,m}|^2}{3 g_n \epsilon_0 \hbar^2} \\
  A_{n,m} &= \frac{\hbar \omega^3}{\pi^2 c^3}B_{n,m},
\end{split}
\end{equation}
where dipole matrix elements connecting ro-vibrational levels, $\mathcal{D}_{n,m}$, are derived in Appendix~\ref{sec:AppHonlLondon}.
\begin{equation} \label{eq:DSummedOverM}
|\boldsymbol{\mathcal{D}}_{m,n}|^2=S_{J_m,J_n}\Big|\int f_{\nu_n}(R) D_e(R) f_{\nu_m}(R)R^2dR\Big|^2.
\end{equation}
The H\"{o}nl-London factors, $S_{J_m,J_n}$, are tabulated in the literature \cite{herzberg,Tatum,Kovacs,whiting} and may be evaluated by the
expressions given in Appendix B. Both the potential energy curve for the molecule and the electronic dipole moment function
$\vect{D}_e^{mol}(R)$ is evaluated with \textit{Gaussian} \cite{gaussian}. From the potential energy curve the ro-vibrational eigenfunctions,
$f_{\nu_n}$, are readily found using the Numerov method, and the one-dimensional integral of Eq.~\eqref{eq:DSummedOverM} can be evaluated. We
use the \textit{Level 7.5} program \cite{LeroyLevel75} to perform these tasks and to evaluate Eq.~\eqref{eq:QMEinstein}, leaving us with the
desired Einstein coefficients.

\begin{figure}[!!th]
  \includegraphics[width=0.42\textwidth]{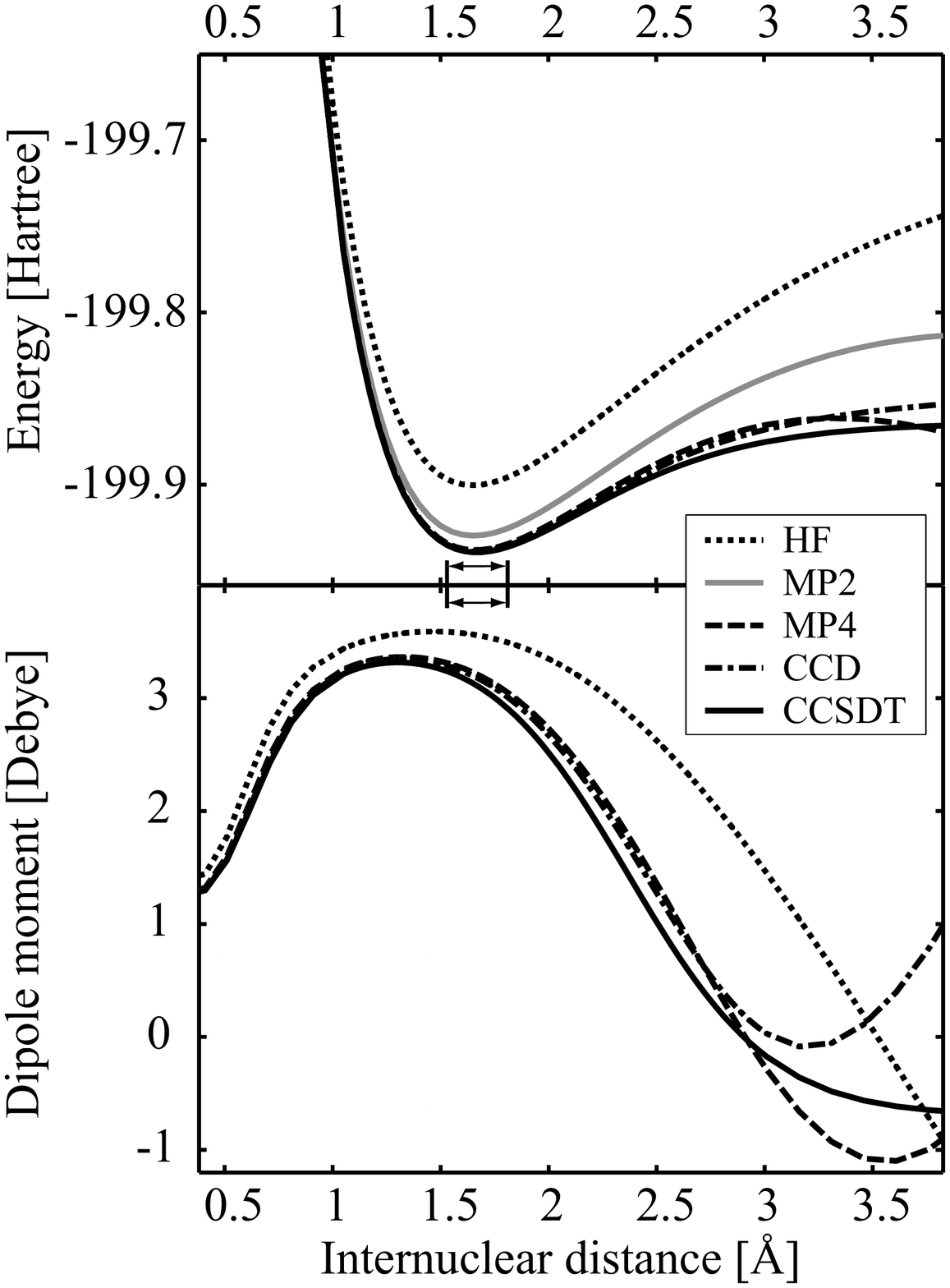}
  \caption{\label{fig:DipoleNPotentialCurveMgH} Top: Born-Oppenheimer electronic potential energy curves of \mgh(X$^1\Sigma$) calculated by
\textit{Gaussian} in a 6-311++G basis set \cite{ExploringChemitryGaussian} using Hartree-Fock (HF) theory, M{\o}ller-Plesset n\emph{th} order
perturbation theory (MPn), coupled cluster theories with singles and double excitations (CCD), and single, double and triple excitations
(CCSDT). See
Refs.~\cite{MoellerPlessetOriginal,MolecularSpectroscopyAbInitioMethods,AdvancesInChemicalPhysics,MoellerPlesset4,CCSDTMethodGaussian} for
descriptions of the methods. The curves for the MP4, CCD and CCSDT calculations are in good agreement close to the equilibrium position,
$1.65$\AA, indicating that these methods give an accurate description of the problem. Bottom: Corresponding dipole moment functions,
$\vect{D}_e^{mol} (R)$ of Eq.\eqref{eq:DefElectronicdipolemoment} pointed along the internuclear axis, of \mgh calculated with
\textit{Gaussian}. The MPn and coupled cluster theories largely agree around the equilibrium distance, although not as well as for the potential
curve due to the dependence on electronic wave functions rather than eigenenergies. The classical turning points for the vibrational ground
state are marked on the common abscissa at $~1.5$ and $1.8$ \AA. The result of the MP2 calculation cannot be discriminated from the MP4 result
at the internuclear distances of interest.
  }
\end{figure}

\subsubsection{Einstein coefficients for \mgh}
Since translational cold samples of \mgh have been produced in a trap loaded with laser cooled Mg$^+$ atomic ions \cite{molhave}, this molecular
ion is the first choice for an implementation of the presented cooling schemes. To our knowledge only a few Einstein coefficients for
transitions within the electronic ground state have been published \cite{vogelius}. We have re-calculated the coefficients using the approach of
the previous section. The potential curves obtained from \textit{Gaussian} \cite{gaussian} using various theoretical approaches on a 6-311++G
basis set \cite{ExploringChemitryGaussian} is given in Fig.~\ref{fig:DipoleNPotentialCurveMgH} together with the corresponding dipole moment
functions in the molecular center of mass system. The potential curves show convincing convergence and our derived vibrational transition
frequencies and equilibrium distance agree with published data within 1.5$\%$ \cite{herzberg}. To compute the accurate electronic dipole moment
function is more challenging, since this requires accurate electronic wave functions. Generally the Møller-Plesset fourth-order perturbation
theory \cite{MoellerPlesset4} and the coupled-cluster theories \cite{CCSDTMethodGaussian} are reliable for the task. The dipole moment functions
converge against a unique function as the level of approximation is refined as shown in Fig.~\ref{fig:DipoleNPotentialCurveMgH} indicating that
the highest order CCSDT function is a good approximation to the physical dipole moment function. Furthermore, we have performed equivalent
calculations on the isoelectronic molecules NaH and BeH${}^+$
\cite{DipoleMomentNaHZemke,BeHDipoleMomentFunction1983,BeHDipoleMomentFunction1991} to compare our results with other published calculations.
The results were in agreement within 5$\%$, a level which is not critical for the simulations of the cooling schemes. The calculated Einstein
coefficients are given in appendix \ref{sec:AppendixEinsteinCoeffs}.

We have now set up the model and acquired the parameters entering the coupling matrix $\vect{K}$ in Eq.~\eqref{eq:RateM} and the solution can
now be found numerically using standard methods as described below.

\subsection{Solving the population dynamics} \label{Sec:SolvingPopulationDynamics}
We model the dynamics of the cooling on the test case of \mgh by solving Eq.~\eqref{eq:RateM} \cite{RungeKuttaRef}. In the population vector,
$\vect{P}$, of Eq.~\eqref{eq:defPopVector} we use $N_{max}=20$ since the population of this and higher-lying levels is effectively zero during
the cooling process. In addition, we omit the $(\nu=2,N)$ states in the cooling schemes if the second excited vibrational state is not coupled
by laser fields. The radiation density $W(\omega)$ at resonance between levels not addressed by lasers has been calculated from a Boltzmann
distribution at $300$ K plus incoherent fields from lamps as described below. The pulsed lasers are included by saturating the pumped
transitions, described in Sec.~\ref{Sec:CoolingSchemeSinglet}, at a repetition rate of $100$ Hz. In the simulation this is done by
redistributing the population in the involved ro-vibrational levels at the given repetition rate according to the degeneracy of the levels. All
simulations are made with populations which are initially Boltzmann distributed at a temperature of 300K. The shape of the incoherent field is
chosen so it maximizes the final population in the rovibrational ground state.

All simulations are made with the most abundant isotopes, in this case $^{24}$Mg$^1$H$^{+}$ (79\%).
\subsection{Efficiency of \singlet cooling schemes} \label{Sec:CoolingEffSinglet}
In Ref.~\cite{vogeliusOptimizeLamp_JPhysB} we found that the optimized radiation density at intermediate timescales induces transitions up to
and including the peak of the population distribution in BBR alone at 300K. Specifically, for \mgh the optimized spectral distribution of the
incoherent source used in the schemes of Fig.~\ref{fig:schemes}(b) and \ref{fig:schemes}(c) is found to be a square distribution with the
maximal density allowed on the rotational transitions  from $(\nu=0,N=1)\leftrightarrow (\nu=0,N=2)$ to $(\nu=0,N=3)\leftrightarrow(\nu=0,N=4)$.
Furthermore, we showed that the spectral radiation density reaching the molecular ions from a realistic lamp is approximately 5 times the
spectral radiation energy density of BBR at 300K. This has been included as a constraint in the optimization of the incoherent field.%
\begin{figure}[!!th]
      \includegraphics[width=0.40\textwidth]{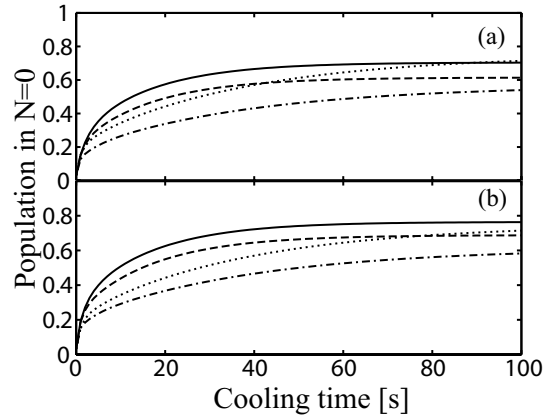}
      \caption{\label{fig:comparedirectraman} Population in the ro-vibrational
        ground state of \mgh(X$^1\Sigma$) vs.~cooling time for the Raman (a) and direct (b) schemes presented in Figs.~\ref{fig:schemes}(b) and
        (c) with the optimized distribution of incoherent radiation (solid), quartz-filtered distribution (dashed) and no incoherent source
        (dash-dot). The same simulation using the scheme of Fig.~\ref{fig:schemes}(a) is depicted for comparison (dot).}
\end{figure}

\begin{figure}[!!th]
  \includegraphics[width=0.42\textwidth]{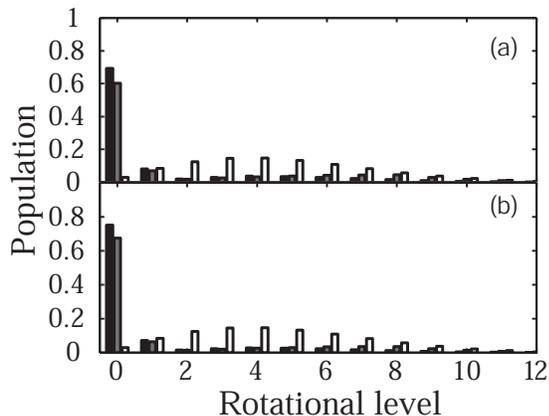}
  \caption{\label{fig:pops} Population distribution in \mgh(X$^1\Sigma$) after 60
    s of cooling using the Raman (a) and direct (b) schemes with the optimized energy distribution of the incoherent source (black), the
    quartz-filtered energy distribution (gray), compared to the initial population distribution at 300 K (white). The ground state population
    after cooling with the optimized incoherent field corresponds to that of a thermal ensemble at $\sim 7$K.}
\end{figure}

While the cooling efficiency at a given time depends critically on the ability to filter out radiation addressing the heating transition in the
low-frequency end of the distribution, it is only weakly depending on the sharpness of the filter in the high-frequency end. This is illustrated
in Figs.~\ref{fig:comparedirectraman} and \ref{fig:pops} by the inclusion of a simulation using a square incoherent field addressing the
transitions $(\nu=0,N=1) \leftrightarrow (\nu=0,N=2)$ to $(\nu=0,N=7) \leftrightarrow (\nu=0,N=8)$ roughly corresponding to the cutoff frequency
of a crystalline quartz window \cite{kimmittbook}. This distribution will be referred to as the "quartz-filtered" distribution below.
Simulations of the evolution of ro-vibrational ground state population of \mgh during cooling with the direct and Raman scheme using these two
incoherent fields are presented in Fig.\ \ref{fig:comparedirectraman} together with the results obtained without the inclusion of an incoherent
source and those obtained by applying the scheme of Ref.~\cite{vogelius} (Fig.~\ref{fig:schemes}(a)).

For very short cooling times no significant difference between the schemes is seen as the relatively slow rotational transitions have not yet
set in. On intermediate time-scales the effect of the added incoherent field is evident and the optimized scheme has an advantage to the scheme
of Ref.\ \cite{vogelius} at times less than $\sim 100$ s. At long times the slower depletion of the $(\nu=0,N=1)$ state using the incoherent
field rather than a laser, as well as the heating effect of the added radiation makes the scheme of Ref.\ \cite{vogelius} more effective than
the other schemes.

The schemes presented here has the advantage of reaching significant cooling after $\sim 30$ s which, combined with the modest demands to
coherent light sources, makes them experimentally attractive. Anticipated performance of traps for neutrals give storage times exceeding 10 s,
comparable to the timescale of the cooling schemes applied on \mgh \cite{meijer1} and in the same regime as \arh, which is a faster candidate
\cite{vogelius}. Hence the application of the schemes may be considered to create rotationally cold neutral molecules in the presence of BBR.

The population distribution after 60 s of cooling is compared with the initial Boltzmann distribution in Fig.\ \ref{fig:pops}. The depletion of
the rotational levels above the $N=2$ ``pump state'' is evident. The difference between using the optimized and quartz-filtered spectral
distribution of incoherent light can be seen in the figures, but the effect is very limited.

The final population in the ro-vibrational ground state of just below 80$\%$, c.f. Fig.~\ref{fig:pops}, corresponds to the ground state
population of a thermal ensemble of \mgh at $\sim$7 K.

\section{Cooling scheme for molecules with rotational sub-structure} \label{sec:rotationalsubstrcture}
In the previous sections, we discussed cooling schemes  applicable to molecular ions with their ro-vibrational energy levels determined by
molecular rotation and vibration only. This will be the case if the relevant electronic state has vanishing total spin and if the projection of
the orbital angular momentum of the electronic state along the internuclear axis is zero, i.e., in \singlet states. We now turn to the other
electronic ground states found in lighter diatomic hydride ions: \doublet, \triplet and \doubletpi

A range of quantum numbers will be needed to describe the rotational sub-states of the molecules to be discussed. We follow the notation of
Herzberg \cite{herzberg} designating the quantum numbers as indicated in Table \ref{tbl:quantumnumbers}. The meaning of the coupled angular
momenta is explained below.

\begin{table}
\caption{\label{tbl:quantumnumbers}Overview of quantum numbers describing the rovibrational state of a molecule, neglecting nuclear spin.
$\hat{\vect{\zeta}}$ denotes a unit vector along the internuclear axis.}
\begin{ruledtabular}
\begin{tabular}{|c|p{7cm}|}
\hline
  Label &  \multicolumn{1}{c|}{Definition} \\
  \hline
  $\vect{L}$ & Total electronic orbital angular momentum \\
  \hline
  $\Lambda$ & Projection of $\vect{L}$ on internuclear axis \\
  \hline
  $\vect{N}$ & Angular momentum of molecular rotation \\
  \hline
  $\vect{S}$ & Total electronic spin \\
  \hline
  $\Sigma$ & Projection of $\vect{S}$ on internuclear axis \\
  \hline
   $M_S$ & Projection of $\vect{S}$ on laboratory Z-axis \\
  \hline
  $\Omega$ & $\Lambda+\Sigma$ \\
  \hline
  $\vect{K}$ & Sum of $\vect{N}$ and $\Lambda\cdot \hat{\vect{\zeta}}$ \\
  \hline
  $\vect{J}$ & Total angular momentum of molecule neglecting nuclear spin \\
  \hline
  $M_J$ & Projection of $\vect{J}$ on laboratory Z-axis \\
   \hline
\end{tabular}
\end{ruledtabular}
\end{table}

We now treat Hunds coupling case (a) and (b) separately and study cooling schemes for both cases.

\subsection{${}^{(2S+1)}\Pi$-states; Hunds case (a)}

An interaction term of the form $H^{so}=A\vect{L}\cdot\vect{S}$ will appear in the Hamiltonian if the projection on the internuclear axis of
both electronic spin, $\vect{S}$, and electronic orbital angular momenta, $\vect{L}$, are nonzero. For moderate rotational excitations this will
normally dominate over terms from the rotational Hamiltonian, $H^{rot}=B\cdot \vect{N}^2$. It is therefore convenient to choose the Hunds case
(a) basis set, consisting of basis functions $\Ket{n,\vect{S}^2\vect{J}^2M_J\Lambda\Sigma\Omega}$ where $n$ is collecting the quantum numbers
defining the molecular state but not mentioned in Table \ref{tbl:quantumnumbers}. In this basis set, the unperturbed Hamiltonian, $H_0$ is
diagonal and the main perturbation term $H^{so}$ is nearly diagonal with the off-diagonal terms satisfying $\Delta\Omega=0$. The
$\Ket{n,\vect{S}^2\vect{J}^2M_J\Lambda\Sigma\Omega}$ basis states are therefore a good approximation to eigenfunctions with good quantum numbers
if $|A| \gg B \cdot J$. In the following section, we restrict the calculation to the pure Hunds case (a) limit where this condition is
fulfilled.
 \doubletpi states are often close to this limit at low rotational excitations and they form the most interesting example of Hunds case (a)
coupling for our purpose, as they are found as ground states of a number of molecules interesting for cooling, including \nh and \hf.

\subsubsection{Energy levels and selection rules}
The first order effect of $H^{so}$ is to split the electronic ground state into states according to the value of $\Omega$. For each of these
states there will be a set of ro-vibrational sub-states arising from $H^{rot}$.

In Hunds case (a), the molecule is well-described as a rotating symmetric top, for which the rotational energies are expressed by
\cite{herzberg}
\begin{equation} \label{eq:pienergies}
F_\nu(J)=B_\nu ( J(J+1)-\Omega^2).
\end{equation}
Here $J$ must take values greater than $|\Omega-N|$ and the lowest rotational state will therefore, in general, have $J\neq0$. The overall
structure of the molecular energy levels can be seen from the sketch of the modified cooling scheme in Fig. \ref{fig:doupletpischeme} for
$S=\frac{1}{2}$.

The case (a) basis state in the laboratory frame can be written as a Wigner rotation of the corresponding wave function in the molecular rest
frame \cite{AmJPhys_angularMomentumStatesofDiatomic_OnHonlLondoncalculation}
\begin{equation} \label{eq:caseaeigenfuncts}
\begin{split}
&\Braket{\{\vect{r}_i\}\vect{R}|nJM_J\Omega S \Sigma} = \\ &\sqrt{\frac{2J+1}{8\pi^2}} \Braket{\{\vect{r}_i^\prime\}
R|n}\Ket{S\Sigma}\mathcal{D}_{M_J\Omega}^{J^\ast}(\alpha \beta \gamma),
\end{split}
\end{equation}
where $\{\vect{r}_i\},\vect{R}$ ($\{\vect{r}_i'\},R$) are the electronic and internuclear coordinates in the laboratory (body-fixed) frame.
Finally $\mathcal{D}_{M\Omega}^{J^\ast}(\alpha \beta \gamma)$ is an element of the Wigner rotation matrix evaluated at the given Euler angles,
$\alpha\beta\gamma$ \cite{BrinkAndSatchlerAngularMomentumBook}. The H\"{o}nl-London factors $S(J^\prime,J^{\prime \prime})$ are found as
outlined in Appendix \ref{sec:AppHonlLondon}
\begin{equation} \label{eq:caseAHonlLondon}
\begin{gathered}
S(J',J^{\prime \prime})=(2J^{\prime \prime}+1) \times \\
|\Braket{J^{\prime\prime}\Omega^{\prime\prime}1 (\Omega^\prime- \Omega^{\prime\prime})| J^\prime \Omega^\prime}|^2
\delta_{S^\prime,S^{\prime\prime}}\delta_{\Sigma^\prime,\Sigma^{\prime\prime}},
\end{gathered}
\end{equation}
where $\Braket{J^{\prime\prime}\Omega^{\prime\prime}1 (\Omega^\prime- \Omega^{\prime\prime})| J^\prime \Omega^\prime}$ is a Clebsch-Gordan
coefficient. This result immediately gives us the following dipole selection rules
\begin{eqnarray} \label{eq:dipoleselectCasea}
\nonumber \Delta J&=&0, \pm1 \quad \textrm{but} \quad J=0 \nleftrightarrow J=0 \\
\Delta \Lambda &=& 0,\pm1, \\
\nonumber \Delta S &=& \Delta \Sigma=0,
\end{eqnarray}
which can also be combined to $\Delta \Omega=0$.

\subsubsection{Cooling schemes}
\begin{figure} [!!!th]
  \includegraphics[width=0.42\textwidth]{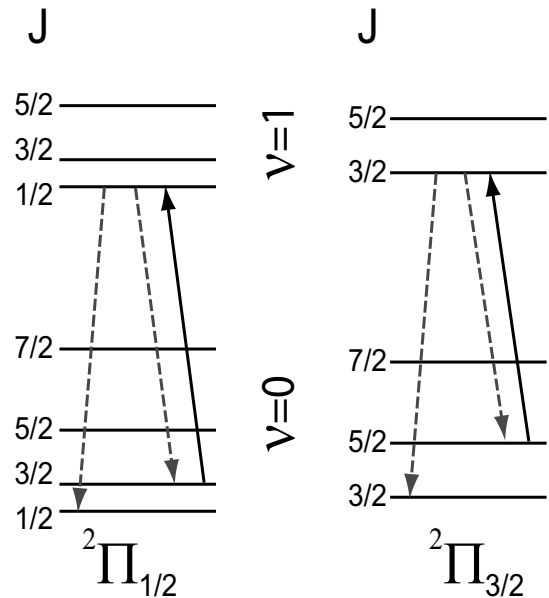}
  \caption{\label{fig:doupletpischeme} Cooling scheme for
    $^{2}\Pi$-states. Each of the two possible values of $\Omega$ results in a series of energy levels and it is necessary to cool the
    $^{2}\Pi_{\frac{1}{2}}$ and $^{2}\Pi_{\frac{3}{2}}$ separately if both are populated. Population is pumped from the first excited rotational
    state in the $\nu=0$ vibrational ground state to the ro-vibrational ground state by a laser induced vibrational
     transition and subsequent spontaneous decays. All the involved transitions are dipole allowed, cf., Eq.~\eqref{eq:dipoleselectCasea}. Solid
lines indicate laser pumped transitions while dashed lines indicate spontaneous decay paths. The $\Lambda$ doubling is not shown in the figure.}
\end{figure}

For \doubletpi molecules we propose the cooling scheme depicted in Fig.~\ref{fig:doupletpischeme}, where we distinguish between the possible
values of $\Omega=\frac{1}{2}$ and $\Omega=\frac{3}{2}$. Since only transitions with $\Delta J = 0,\pm1$ are allowed, we can pump population
from the first excited rotational state in the vibrational ground state to the rotational ground state of the first excited vibrational level.
The former is denoted the "pump state" in analogy with the nomenclature in Sec.~\ref{Sec:CoolingSchemeSinglet}. From the $(\nu=1,J=\Omega)$
state spontaneous emission brings population either back to the pump state or down to the ro-vibrational ground state. The cooling scheme must
be applied for each populated $\Omega$ state individually. In Fig.\ \ref{fig:doupletpischeme} we have assumed population of both
$\Omega=\frac{1}{2}$ and $\Omega=\frac{3}{2}$. In the absence of incoherent radiation this forms a pumping cycle where population initially in
the $(\nu,J)=(0,\Omega+1)$ state is transferred to the ro-vibrational ground state. As in the singlet case, the presence of BBR and possibly
additional incoherent radiation from a lamp, will induce rotational transitions and thereby feed the pump state with population from
higher-lying states. The entire population is therefore cooled.

Cooling schemes for other Hunds case (a) molecules may be derived from straightforward generalization of the \doubletpi scheme.

\subsubsection{Numerical simulations} \label{sec:NumericalSimRotationalSubstructureCasea} The simulation is done using the approach described in
Sec.~\ref{Sec:SolvingPopulationDynamics} but with the dipole transition matrix elements calculated using the H\"{o}nl-London factors of
Eq.~\eqref{eq:caseAHonlLondon}. We have chosen the molecule \hf as an example of a \doubletpi ground state molecule.

Since the spin-orbit coupling parameter $A=-292$ cm$^{-1}$ is much larger in magnitude than the rotational constant $B=17$ cm$^{-1}$ \hf is best
described in the Hunds case (a) scheme \cite{HF_and_HCl_SpectroscopicData}. The appropriate cooling scheme is depicted in
Fig.~\ref{fig:doupletpischeme}, although it should be noted that, for \hf, $\Omega=\frac{3}{2}$ is the lower state. To model the cooling scheme
we use the dipole moment functions in Ref.~\cite{HF_and_HCl_DipoleAndEinstein} and the accurate spectroscopic data of
Ref.~\cite{HF_and_HCl_SpectroscopicData}.

\begin{figure}[!!t]
  \includegraphics[width=0.42\textwidth]{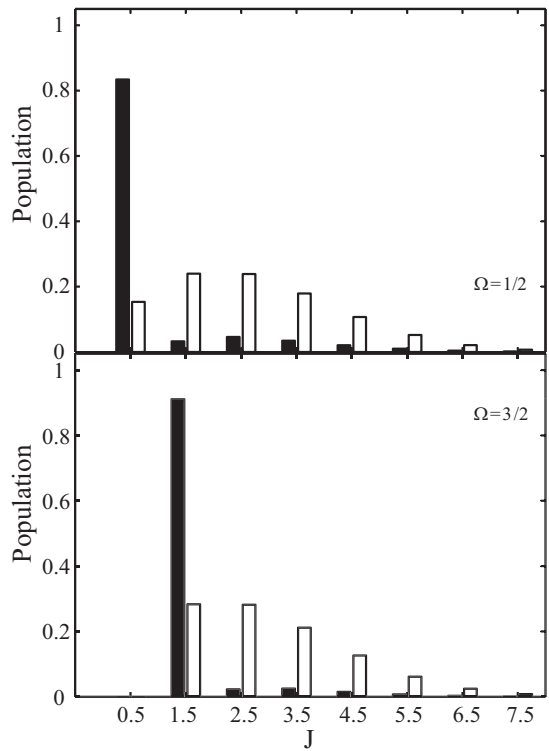}
  \caption{\label{fig:Coolingdoubletpibars} Cooling efficiency of \hf(X$^2\Pi$) in the two $\Omega$ sub-states. In the dipole approximation the
two sub-states are uncoupled if pure Hunds case (a) applies. The cooling scheme will therefore be significantly simplified if one can design the
experiment such that only the lowest $\Omega=\frac{3}{2}$ state is populated in the cooling scheme}
\end{figure}

\begin{figure}[!!t]
      \includegraphics[width=0.45\textwidth]{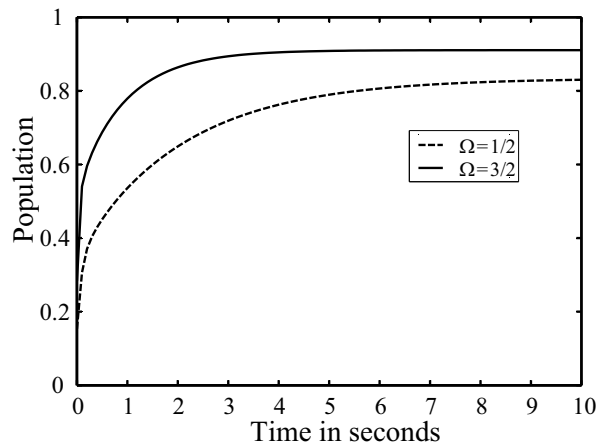}
      \caption{\label{fig:CoolingdoubletpiTime} Cooling efficiency as function of cooling time for the two $\Omega$ sub-states of the \doubletpi
electronic ground state of \hf. The cooling is seen to reach steady-state after $\lesssim10$ s without the inclusion of an incoherent source,
largely due to the large permanent dipole moment of \hf.}
  \end{figure}

In the cooling scheme of Fig.~\ref{fig:doupletpischeme} the pumping is done from the first excited rotational level. This fact, combined with a
large permanent dipole moment and hence rotational transition rate of \hf (2.57 Debye), makes the effect of the broadband incoherent radiation
marginal. We have therefore performed the simulations without the inclusion of an incoherent source. The results of simulations are given for
both the \doubletpi$_{1/2}$ and \doubletpi$_{3/2}$ states in Figs.~\ref{fig:Coolingdoubletpibars} and \ref{fig:CoolingdoubletpiTime}.

Further splitting of the levels indicated in Fig.\ \ref{fig:doupletpischeme} will appear due to $\Lambda$ doubling. The effect is largest in the
\doubletpi$_{\frac{1}{2}}$ state where it has a magnitude on the order of 10 GHz, which is more than one can expect to cover with the bandwidth
of a single pulsed laser. Therefore the laser transitions indicated for the \doubletpi$_\frac{1}{2}$ scheme needs to be divided into two. The
splitting of the lowest \doubletpi$_{\frac{3}{2}}$ state is an order of magnitude smaller, so it is not necessary to split that laser transition
if a pulsed laser system is used. This leaves us with three laser frequencies to use for the cooling scheme if we assume that both
$\Omega=\frac{1}{2}$ and $\Omega=\frac{3}{2}$ are populated.

Complications arise if we are not in the pure Hunds case (a) scheme. This occurs if the rotational part of the Hamiltonian cannot be neglected
compared to the spin-orbit part. Treated in the case (a) basis, the rotational part will produce non diagonal perturbations
\cite{Lefebvre-BrionGrundigAngularMomentum}. This would allow a coupling from $(\nu=1,J=\Omega)\rightarrow(\nu=0,J=\Omega+2)$ (the introduction
of quadrupole couplings would have a similar effect). We do not expect this effect to be significant given the difference between $|A|$ and $B$.
We did, however, check the stability of the scheme when introducing such couplings and found that due to the fast rotational redistribution
rates, the population that was coupled out of the cooling cycle by $\Delta J=2$ transitions would rapidly be taken back. The negative effect of
such couplings is small (less than 10\% decrease in cooling efficiency) if the $\Delta J=\pm2$ couplings are less than 20\% of the $\Delta J =0$
coupling strength.

It should be noted that since the coupling between the $\Omega$ states is also absent in the pure case (a) coupling, it would be possible to
prepare the sample so that only the $\Omega=\frac{3}{2}$ sub-state is populated, due to its significantly lower energy. This would make the
lasers addressing the other level superfluous. In that case only a single laser frequency is needed to cool the molecules.

\subsection{$^{(2S+1)}\Sigma$-states; Hunds case (b)} \label{sec:HundsCaseBsubsec}

If $B\gtrsim|A|$ or at high rotational excitations the Hunds case (a) basis functions will no longer be approximate energy eigenfunctions. If
$H^{rot}$ dominates, the Hunds case (b) basis, $\Ket{n\vect{J}^2M_J\vect{N}^2\vect{S}^2\Lambda}$ is convenient as the total Hamiltonian is
nearly diagonal in this basis. In particular this is fulfilled for $^{2S+1}\Sigma$ states which are common as electronic ground states of light
diatomic molecular ions, including \bh (X$^2\Sigma$) and \oh (X$^3\Sigma$).  Below we treat the \doublet and \triplet cases separately.

\subsubsection{Energy levels of doublet states}

The sub-states of a rotational level in a  molecule in a \doublet state are split due to the interaction of the spin of the unpaired electron
and the molecular rotational angular momentum. This is due to the spin-rotation Hamiltonian, $H^{sr}=\gamma \vect{N}\cdot\vect{S}$, with
$\gamma$ denoting the spin-rotation coupling constant. The resulting energies of the doublet are given by \cite{herzberg}
\begin{eqnarray} \label{eq:doubletsplit}
    F_1(N)&=& B N(N+1)+\frac{1}{2}\gamma N, \\
    F_2(N)&=& B N(N+1)-\frac{1}{2}\gamma (N+1),
\end{eqnarray}
and the sub-states are denoted $F_1$ and $F_2$ for $J=N+\frac{1}{2}$ and $J=N-\frac{1}{2}$ respectively.

\subsubsection{Energy levels of triplet states}
Molecular ions in \triplet electronic states will, apart from the spin-rotation splitting discussed above, have an additional splitting from the
coupling of the electronic spin of the two unpaired electrons. Such states are relatively rare, as pairing of the electronic spins is usually
favored. Nevertheless, the ionic hydrides in the 16\textit{th} group of the periodic table, including \oh and \sh, have such electronic ground
states and we therefore consider the applicability of the cooling schemes to such states here. The energies of the three spin sub-states are
given by \cite{herzberg}
\begin{eqnarray} \label{eq:tripletsplit}
    F_1(N)&=& B N(N+1)+\frac{2\lambda(N+1)}{2N+3}+\gamma (N+1), \\
    F_2(N)&=& B N(N+1), \\
    F_3(N)&=& B N(N+1)-\frac{2\lambda N}{2N-1}-\gamma N.
\end{eqnarray}
In analogy with the doublet case $F_1$, $F_2$, and $F_3$ denotes the sub-states with $J=N+1$, $J=N$ and $J=N-1$, respectively. In the expression
$\gamma$ is the spin-rotation coupling constant and $\lambda$ is the spin-spin--splitting constant. The latter is normally an order of magnitude
or more larger than $\gamma$ and, consequently, the multiplet splitting of triplet states at moderate rotational excitations are much greater
than the corresponding splittings of a doublet electronic states.

\subsubsection{Selection rules} \label{sec:CaseBSelcectRules}
\begin{figure} [!!t]
  \includegraphics[width=0.37\textwidth]{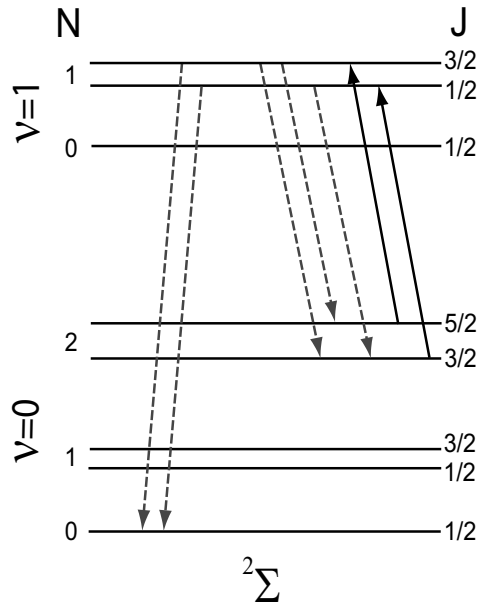}
  \caption{\label{fig:doupletscheme} Cooling scheme for
    $^{2}\Sigma$-states. Due to the spin-rotation coupling each rotational quantum state $N$ split into two sub-levels with
    $J=|N+\frac{1}{2}|,|N-\frac{1}{2}|$. The dipole-allowed vibrational transitions are indicated on the figure using solid lines for laser
    pumped transitions and dashed lines for subsequent spontaneous decay paths.}
\end{figure}

\begin{figure} [!!h]
  \includegraphics[width=0.30\textwidth]{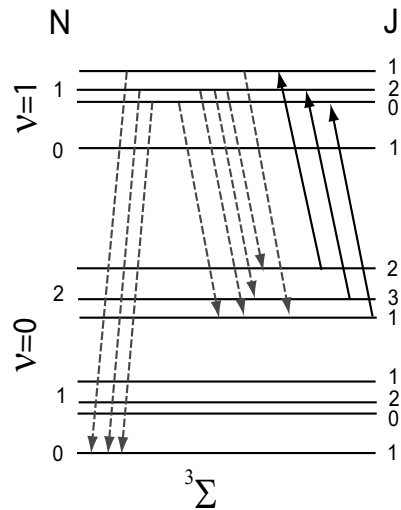}
  \caption{\label{fig:tripletscheme} Cooling scheme for
    $^{3}\Sigma$-states. Due to spin-spin and spin-rotation coupling each rotational quantum state $N$ split into sub-levels with
    $J=|N+1|,N,|N-1|$. The dipole-allowed vibrational transitions are included in the figure with solid lines to indicate laser pumped
    transitions and dashed lines to indicate spontaneous decay paths.}
\end{figure}

In Hunds case (b) the good quantum numbers are $N,S,J,M_J$ and $\Lambda$. We therefore write the eigenfunctions in the laboratory frame as
\begin{equation}
\begin{gathered} \label{eq:caseBeigenfuncts}
\Braket{\{\vect{r}_i\},\vect{R}|nJM_J N S,\Lambda}=\sqrt{\frac{2N+1}{8\pi^2}} \times \\ \sum_{M_S=-S}^S \sum_{M_N=-N}^N
\Braket{\{\vect{r}_i^\prime\},R|n}\Braket{N M_N S M_S| J M_J} \times \\
\Ket{S M_S} \mathcal{D}_{M_N\Lambda}^{N^\ast}(\alpha \beta \gamma),
\end{gathered}
\end{equation}
where $\{\vect{r}_i\},\vect{R}$ ($\{\vect{r}_i'\},R$) are the electronic and internuclear coordinates in the laboratory (body-fixed) frame. We
then follow the approach of Appendix \ref{sec:AppHonlLondon} to get the H\"{o}nl-London factors
\begin{equation} \label{caseBHonlLondon}
\begin{gathered}
S(J',J'')=(2N''+1)(2J'+1)(2J''+1)\times \\
\Braket{N'' \Lambda'' 1 (\Lambda'-\Lambda'')|N' \Lambda'}^2  \begin{Bmatrix} S' & N'' & J''\\ 1 & J' & N' \end{Bmatrix}^2 \delta_{S'S''},
\end{gathered}
\end{equation}
where $\begin{Bmatrix} S' & N'' & J''\\ 1 & J' & N' \end{Bmatrix} $ is a 6j symbol \cite{BrinkAndSatchlerAngularMomentumBook}. The following
selection rules are extracted
\begin{equation}
\begin{split} \label{eq:dipoleselectCaseb1}
\Delta J&=0, \pm1 \quad \textrm{but} \quad J=0 \nleftrightarrow J=0 \\
\Delta N&=0,\pm1 \quad \textrm{but }\ \Delta N \neq 0 \quad \textrm{if $\Lambda'=\Lambda''=0$} \\
\Delta \Lambda &= 0,\pm1, \\
\Delta S &=0.
\end{split}
\end{equation}

\subsubsection{cooling scheme} The cooling scheme proposed for Hunds case (b) molecules closely resembles the singlet cooling scheme. It is
depicted in Fig.~\ref{fig:doupletscheme} for \doublet molecules and in Fig.~\ref{fig:tripletscheme} for \triplet molecules. The optical pumping
is done from the $(\nu=0,N=2,J)$ set of states to $(\nu=1,N=1,J')$. Then dipole allowed spontaneous decay will result in transitions back to the
"pump states" or to the non-degenerate ro-vibrational ground state. The only change to the scheme when compared to the singlet case is to assure
the addressing of all sub-states in the $N$ multiplet. This is possible because the $\Delta N=\pm 1$ selection rule for $\Sigma$ states from
Eq.~\eqref{eq:dipoleselectCaseb1} is the same as in the singlet case. The role of BBR and additional incoherent radiation is the same as in the
previous schemes.

The number of transitions to be pumped is three for the \triplet states and two for the \doublet states. The splitting of the levels in the
former is expected to be much larger than for the \doublet case, since the spin-spin coupling parameter $\lambda$ is much greater than the
spin-rotation parameter $\gamma$ as mentioned in Sec.~\ref{sec:HundsCaseBsubsec}.

\begin{figure}[H]
      \includegraphics[width=0.45\textwidth]{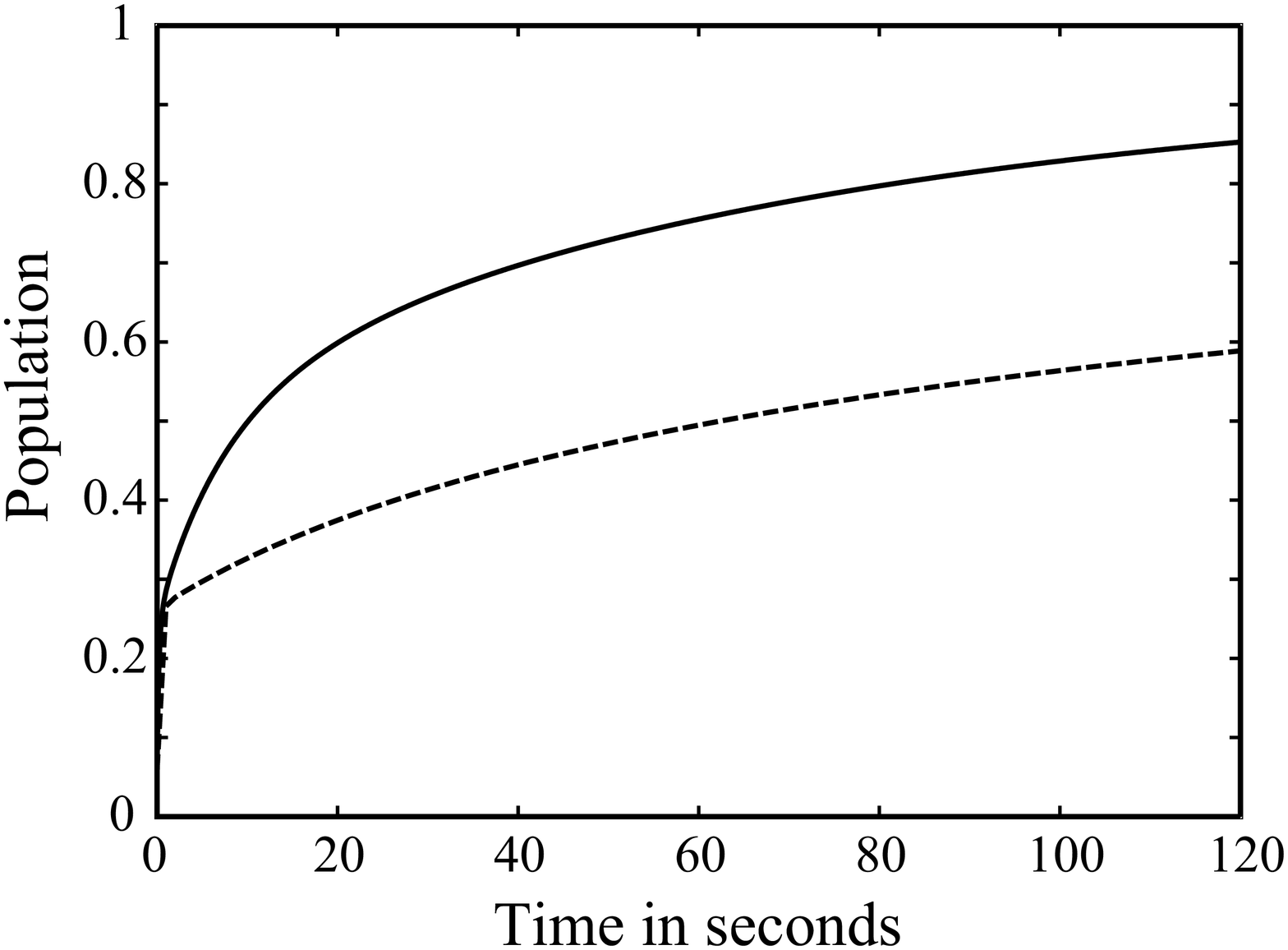}
      \caption{\label{fig:BHtime120s} Population in the lowest rotational state of \bh(X$^2\Sigma$) as function of cooling time. Simulations are
made using the scheme of Fig.\ \ref{fig:doupletscheme} with BBR only (dashed line) and with the inclusion of the field from an incoherent source
addressing the $N=1\rightarrow N=2$ and $N=2 \rightarrow N=3$ transitions (solid line). We see that a significant improvement is obtainable
using the incoherent source. In line with our experience from \mgh there is only a couple of percent loss of cooling efficiency when using a
softer low-frequency pass filter, for example, letting the broadband incoherent radiation extend to include transitions up to and including $N=4
\rightarrow N=5$.}
\end{figure}

\begin{figure}[!!b]
      \includegraphics[width=0.45\textwidth]{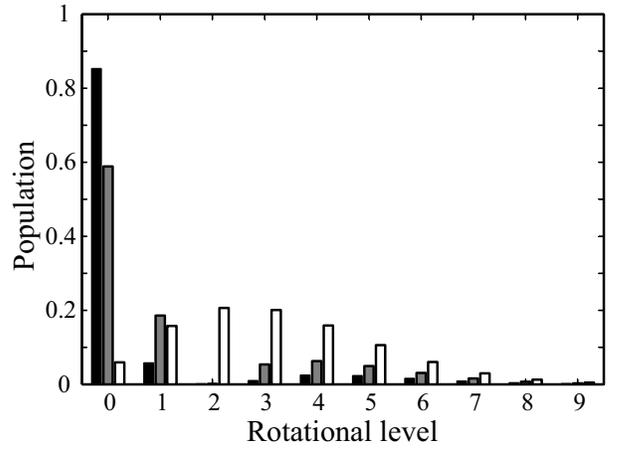}
      \caption{\label{fig:BHbars120s} Population in the lowest rotational states of \bh(X$^2\Sigma$) after cooling in 120 s using the incoherent
      radiation from a lamp addressing the $N=1\rightarrow N=2$ and $N=2 \rightarrow N=3$ transition (Black columns) and in BBR only (grey
      columns). The initial 300K Boltzmann distribution is included for comparison (unfilled columns). We note that slightly better cooling
      efficiency should can be obtained using longer cooling times, cf Fig. \ref{fig:BHtime120s}. This is, however, impractical and the
      obtainable improvements would be rather small. The substructure of the rotational levels is included in the simulation but omitted on the
      figure.}
  \end{figure}

\subsubsection{Numerical simulations; \bh (\doublet)} \label{sec:NumericalSimRotationalSubstructureCaseb}

 Here we treat \bh as an example of a \doublet ground state molecule
and discuss the molecule specific parameters and their implications on the cooling schemes.

The numerical simulation is done for $^{11}$B$^{1}$H$^{+}$ which is the dominant isotope (80\%). We use the potential energy and dipole moment
functions of Ref.~\cite{BHandALH_Dipole_Potential}. With those functions, we use the approach of Sec.\ \ref{sec:CaseBSelcectRules} to calculate
the matrix of Einstein coefficients between rotational and vibrational states. Finally, we make the simulation as described in Sec.\
\ref{Sec:SolvingPopulationDynamics} but with the modified energy level structure. If one neglects fine-structure, the laser wavelength for the
two, then identical, pump transitions depicted in Fig.\ \ref{fig:doupletscheme}, is $\Omega_0=4.17 \mu$m. The real resonant transition
frequencies are shifted from this central frequency through Eq.\ \eqref{eq:doubletsplit} where $\gamma=-0.014$ cm$^{-1}$ \cite{herzberg}. This
gives a splitting of laser frequencies, including fine structure, of $\pm0.007 \:$cm$^{-1}\simeq \pm 210\: $MHz. This difference is comfortably
smaller than the typical bandwidth of a pulsed laser system. The hyperfine coupling coefficient has, to our knowledge, not been calculated.
Typical values are, however, on the order tens to hundreds of MHz, allowing us to address all hyperfine substates with the same pulsed laser
system. Hence, it is reasonable to expect that for practical implementations only a single, pulsed laser frequency is needed.

The results of a numerical simulation are given in Figs.\ \ref{fig:BHtime120s} and \ref{fig:BHbars120s}. We note that the convergence is quite
slow compared to what we saw from \mgh and \hf. Optimal cooling is not obtained until after $\sim 2$ minutes. This is not too critical as 60 \%
of the population is in the ground state after 20 s. As expected from the discussion in Ref.~\cite{vogeliusOptimizeLamp_JPhysB} we find, that
the optimized distribution of the incoherent source addresses the transitions $(\nu=0,N=1)\leftrightarrow(\nu=0,N=2)$ and
$(\nu=0,N=2)\leftrightarrow(\nu=0,N=3)$. Similarly it is confirmed, that the cooling efficiency has little sensitivity towards the high
frequency cutoff of the incoherent field.

\subsubsection{Numerical simulations; \oh (\triplet)}

\begin{figure}[!!thb]
  \includegraphics[width=0.45\textwidth]{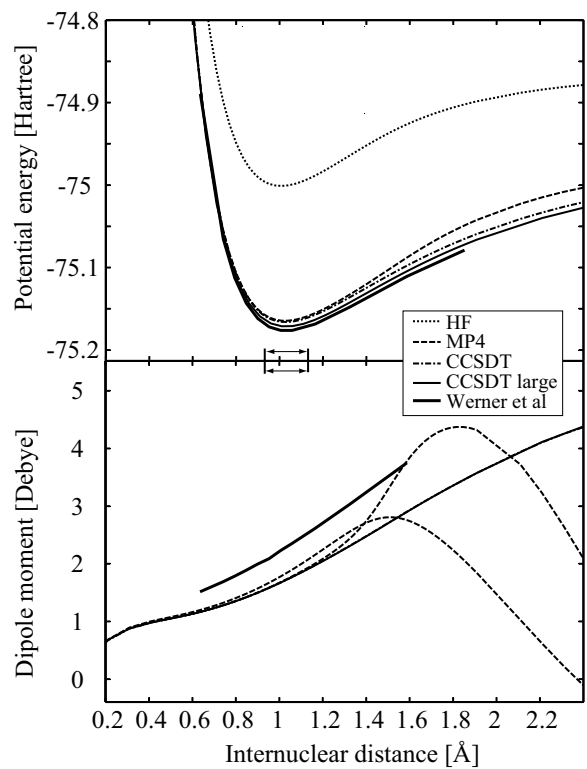}
  \caption{\label{fig:DipoleNPotentialCurveOH} Top: Born-Oppenheimer potential curves for X$^3 \Sigma$ \oh calculated by \textit{Gaussian} with
various theoretical models compared to the calculation of Ref.~\cite{dipoleOH_Werner}. All our calculations are done in a 6-311++G basis set
except the solid black line which is made in the generally more accurate aug-cc-pVTZ basis \cite{ExploringChemitryGaussian}. The curves agree
close to the equilibrium, $1.03$\AA, for the MP4 (Fourth order M{\o}ller-Plesset perturbation theory) and Coupled Cluster approaches (CCD,
CCSDT) indicating an accurate level of theory. The different methods are described in Refs.
\cite{MoellerPlessetOriginal,MolecularSpectroscopyAbInitioMethods,AdvancesInChemicalPhysics,MoellerPlesset4,CCSDTMethodGaussian}. Bottom: Dipole
moment function calculated with \textit{Gaussian} using similar levels of theory and basis sets. The agreement between the calculations is
reasonable and the effect of using the larger basis set for the CCSDT theory is not visible on the given scale, but our results show some
discrepancy with the results of Ref.~\cite{dipoleOH_Werner}. This small discrepancy, however, has very little effect on the cooling scheme. The
classical turning points for the vibrational ground state are marked on the common abscissa at $0.95$ and $1.15$ \AA . The dipole moment
functions are given in center of mass coordinates.
  }
\end{figure}

As an example of a molecule with the \triplet ground state we have chosen \oh. This molecule plays an important role in the chemistry in comet
tails \cite{ComettailOh}, the upper earth atmosphere and interstellar clouds \cite{DRTheoryExpApplications}. The electronic ground state of \oh
is $^3\Sigma^-$. The effect of hyperfine splittings is expected to be much smaller than the bandwidth of a typical pulsed laser system due to
the nuclear spins $I=0$ and $I=\frac{1}{2}$ of O and H respectively. Hence the molecule is well-described by the level scheme of
Fig.~\ref{fig:tripletscheme}. The frequencies of the three laser beams required are found from Eq.\ \eqref{eq:tripletsplit} and the constants
$\gamma=-0.0147$ cm$^{-1}$ and $\lambda=2.13$ cm$^{-1}$ \cite{herzberg}. The wavelength of the un-split transition is $3.3$ $\mu$m with the
three sub-transitions shifted -3.2 GHz, 0 and -12 GHz with respect to it. This splitting is to large to be covered by a single broad laser
unless one finds a way to generate shorter and thereby broader and more intense pulses in this wavelength regime. This is an obvious
experimental complication that will often arise in the case of \triplet states due to the generally large value of the spin-spin splitting
constant $\lambda$. It should, however, be noted that the 3 GHz may be covered by a single pulsed laser, leaving only two laser frequencies in
the cooling scheme. We have calculated the dipole moment functions of \oh and compared our results to Ref.~\cite{dipoleOH_Werner} in
Fig.~\ref{fig:DipoleNPotentialCurveOH}. In the simulations we use the function obtained in the CCSDT (aug-cc-pVTZ) calculation.

\begin{figure}[!!t]
      \includegraphics[width=0.45\textwidth]{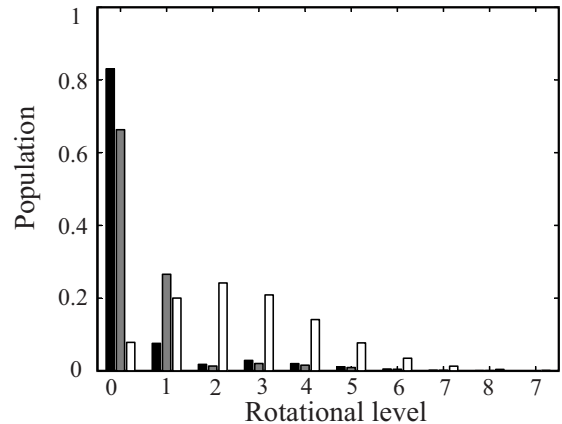}
      \caption{\label{fig:OHpopulation10s} Population in the lowest rotational states of \oh(X$^3\Sigma$) after cooling in 10 s using the
       incoherent radiation from a lamp addressing the $N=1\rightarrow N=2$ and $N=2 \rightarrow N=3$ transition (Black columns) and in BBR only
      (grey columns). The initial 300K Boltzmann distribution is included for comparison (unfilled columns). The spin substructure of the rotational levels
      is included in the simulation but omitted on the figure.
      }
  \end{figure}

The final population distribution in the numerical simulation is given in Fig.\ \ref{fig:OHpopulation10s}. The scheme is both faster and more
effective than what was found for \mgh. This can be understood from comparison of the Figs.~\ref{fig:DipoleNPotentialCurveMgH} and
\ref{fig:DipoleNPotentialCurveOH}. A larger gradient of the dipole moment function of \oh results in a higher effective pump rate from $N=2$ to
$N=0$. As with \mgh we see a significant increase in the cooling efficiency when introducing broadband radiation from an incoherent source to
deplete the $N=1$ population.

The simulation shows the efficiency of the rotational redistribution in the \triplet state. Considering the nonzero line strengths for
transitions between the $F_i,F_J$,  $(i\neq j)$ series of states, provided $\Delta N=\pm1$, one could be tempted to omit one or more laser
frequencies expecting rotational redistribution to empty the remaining substates by rotational transitions through neighboring $N$-levels.
Unfortunately such redistribution rates, requiring two or more rotational transitions through specific substates, are much too slow to have a
significant effect on the cooling scheme. Therefore each of the three laser frequencies are needed to make the cooling scheme effective. In
accordance with the previous results we find that the optimized distribution of incoherent radiation from a lamp addresses only the
$(\nu=0,N=1)\leftrightarrow(\nu=0,N=2)$ transition.

Finally, it should be noted that $^{2S+1}\Sigma$-states are always cases of pure case (b) coupling due to the vanishing orbital angular momentum
and the selection rule in $N$ is close to exact. This stands in contrast to $^{2S+1}\Pi$ states which often have effects of intermediate
coupling which will complicate the suggested case (a) cooling scheme further.

\section{Summary} \label{sec:summary}
We have presented cooling schemes for rotational cooling of translational cold molecular ions in the \singlet, \doublet, \triplet and \doubletpi
electronic ground states. For all but the relatively rare \triplet electronic state the schemes can be realized by optical pumping with a single
pulsed laser beam, possibly combined with the inclusion of a broad-band incoherent source. They are therefore experimentally attractive, and
preliminary experiments are presently under way with \mgh.

Possible applications include high-precision spectroscopy and measurements of absolute reaction rates with molecular ions in a single, well
defined quantum state. This could for example be used to study dissociative recombination with unprecedented resolution or molecular reactions
in interstellar media or comet tails \cite{IAUsymposiumProceedStellarMolecIons,IAUsymposiumProceedCometsMolecIons}. Ultimately, the access to
cold molecular ions could be used in implementations of quantum logics.

\begin{acknowledgments}
L. B. M. is supported by the Danish Natural Science Research Council (Grant no. 21-03-0163). M.D. acknowledges financial support from the Danish
National Research Foundation through the Quantum Optics Center QUANTOP.
\end{acknowledgments}
\pagebreak
\appendix
\section{Einstein coefficients} \label{sec:AppendixEinsteinCoeffs}
\onecolumngrid

\begin{table}[!!h]
\caption{Einstein coefficients for selected transitions in s$^{-1}$ and corresponding transition frequencies in cm$^{-1}$. In the table the
quantities have the following meaning: $A^{rot}=A(\nu=0,N=1\rightarrow\nu=0,N=0)$, $A^{vib}=A(\nu=1,N=1\rightarrow\nu=0,N=0)$ for the
$\Sigma$-states and $A^{rot}=A(\nu=0,J=\Omega+1\rightarrow\nu=0,J=\Omega)$, $A^{vib}=A(\nu=1,J=\Omega\rightarrow\nu=0,J=\Omega)$ for the
$\Pi$-state. A similar notation is used for the transition frequencies. The pure rotational transition rates (first column) indicate the
rotational redistribution speed while the vibrational transition rates give the spontaneous decay rate from the excited vibrational state in the
pumping schemes. The data largely explains the qualitative difference in cooling efficiency for the molecular ions. Large rotational
redistribution rates indicate a fast scheme, while fast spontaneous decays from the excited vibrational state indicate high effective pump rate,
i.e. high cooling efficiency. The data for \bh is found using the data from Ref.~\cite{BHandALH_Dipole_Potential} and the computer programme of
Ref.~\cite{LeroyLevel75} from which the H\"{o}nl-London factors were corrected to conform with the multiplet expressions of
Sec.~\ref{sec:rotationalsubstrcture}. Data for \oh was found using the same approach and data from Fig.~\ref{fig:DipoleNPotentialCurveOH}.
Finally the data on \hf was obtained using the data of Refs.~\cite{HF_and_HCl_SpectroscopicData,HF_and_HCl_DipoleAndEinstein}.}
\begin{ruledtabular}
\begin{tabular}{|c|c|c|c|c|}
   & A$^{rot}$ (S$^{-1}$) & A$^{vib}$ (S$^{-1}$) & $\omega^{rot}$ (cm$^{-1}$)& $\omega^{vib}$ (cm$^{-1}$)\\ \hline

  $^{24}$Mg$^1$H$^+$ (X\singlet)  & $2.5 \cdot 10^{-3}$  & $20.5$  & 12.9 & 1672 \\

  \hline $^{11}$B$^1$H$^+$ (X\doublet)  & $0.2 \cdot 10^{-3}$ (Q-branch)  & 11.5 & 25.0 & 2437 \\
     & $0.4 \cdot 10^{-3}$ (R-branch) & 23.0 &  &  \\  \hline
   $^{16}$O$^1$H$^+$ (X\triplet)  & $3.8 \cdot 10^{-3}$ (P-branch) & 18.3 &  &  \\
     & $19.2 \cdot 10^{-3}$ (R-branch) & 91.6 & 33.07 & 2990 \\
    & $11.5 \cdot 10^{-3}$ (Q-branch) & 54.9 & & \\ \hline \hline

    & A$^{rot}$ (S$^{-1}$)& A$^{vib}$ (S$^{-1}$)& $\omega^{rot}$ (cm$^{-1}$)& $\omega^{vib}$ (cm$^{-1}$) \\
    \hline

  $^{19}$F$^1$H$^+$ (X\doubletpi) & $ 93.8 \cdot 10^{-3}$ ($\Omega=\frac{1}{2}$) & 82.4 & 51.6 & 2964 \\

  & $ 347 \cdot 10^{-3}$ ($\Omega=\frac{3}{2}$) & 98.9 & 85.5 & 2999\\ \hline

\end{tabular}
\end{ruledtabular}
\end{table}

\twocolumngrid

\section{H\"{o}nl-London factors} \label{sec:AppHonlLondon}
For completeness we include the details of the derivation of Eqs.~\eqref{eq:DSummedOverM},\eqref{eq:caseAHonlLondon} and
\eqref{caseBHonlLondon}.

\subsection{\singlet ground state}

Eq. \eqref{eq:QMEinsteinNonDegenerate} must be modified if $\Psi_m$ and $\Psi_n$ are degenerate. The effective Einstein B-coefficient is found
as $B_{n,m}=\sum_{\mu} \sum_{\xi} \frac{B_{m_\xi,n_\mu}}{g_n}$, where $\mu,\xi$ denote the sub-states of $\Psi_n$ and $\Psi_m$, respectively,
and $g_n$ the degeneracy of the initial (upper) state. One summation is done to include transitions to all sub-states of the final state, $m$,
while the remaining terms correspond to averaging the result over the sub-states of the initial state. We then define the total transition
dipole moment for degenerate states as
\begin{equation} \label{eq:DefDipoleSummed}
|\mathcal{D}_{n,m}|^2=\sum_{\xi,\mu} |D_{n_\xi,m_\mu}|^2,
\end{equation}
where the summation is done over all transitions between sub-states of the system. The Einstein coefficients between degenerate states then take
the form
\begin{equation} \label{eq:AppQMEinstein}
\begin{split}
  B_{n,m} &= \frac{\pi |\mathcal{D}_{n,m}|^2}{3 g_n \epsilon_0 \hbar^2} \\
  A_{n,m} &= \frac{\hbar \omega^3}{\pi^2 c^3}B_{n,m},
\end{split}
\end{equation}
Which is the same as Eq.~\eqref{eq:QMEinsteinNonDegenerate} except for the degeneracy factor. We now move to a molecule-fixed coordinate system.
We define the electronic dipole moment function by integrating the dipole operator, $\vect{M}^{mol}$, over the electronic variables, $\tau_e$
\begin{equation} \label{eq:DefElectronicdipolemoment}
\vect{D}_e^{mol} (R)=\int \psi_e (\{\vect{r}_i'\},R)^{\ast} \vect{M}^{mol} \psi_e (\{\vect{r}_i'\},R) d \tau_e.
\end{equation}
Here we stay in the electronic state defined by the wave function $\psi_e(R)$ in the body-fixed frame. We have calculated $\vect{D}_e^{mol} (R)$
\textit{ab initio} with \textit{Gaussian} \cite{gaussian}. Details of these calculations are molecule-specific and will be given below.

To transform the dipole moment to the laboratory frame we now specialize to \singlet states, postponing the general solution to
Sec.~\ref{sec:rotationalsubstrcture}. For \singlet diatomic molecules the cylindrical symmetry of the potential will ensure that
$\vect{D}_e^{mol}(R)$ points along the internuclear axis. Hence the $Z$-component of $\vect{D}_e^{mol} (R)$ in the laboratory system is given by
\begin{equation} \label{eq:electronicDipoleZaxis}
\vect{D}_e^{lab}(R)_Z=\vect{D}^{mol}_e(R)\cdot\hat{Z}=D_e^{mol}(R) \cos \theta.
\end{equation}

The molecular states are degenerate so it is necessary to sum over all sub-states to obtain the transition dipole moment defined in
Eq.~\eqref{eq:DefDipoleSummed}. In the case of \singlet molecules this corresponds to summing over all projections $M_J$ of the molecular
angular momentum $\vect{J}$. In carrying out the summation over the sub-states in Eq.~\eqref{eq:DefDipoleSummed} the selection rule $\Delta
M_J=0,\pm 1$ makes it possible to rewrite the expression as a single sum over $\mu=M_J$ which can be related to the total transition dipole
moment. Since the transition probability must be independent of the orientation of the laboratory coordinate system we have
\begin{equation} \label{eq:DSumSubstates}
|\boldsymbol{\mathcal{D}}_{m,n}^{lab}|^2=\sum_{M_J} |\vect{D}^{lab}|^2=3\sum_{M_J}|\vect{D}_Z^{lab}|^2.
\end{equation}

Inserting Eq.~\eqref{eq:electronicDipoleZaxis} in Eq.~\eqref{eq:Defdipolemoment} we find

\begin{equation}
\begin{split} \label{eq:DipolemomentZaxis}
\vect{D}^{lab}_Z=\int & \psi_{r_n,\nu_n}^{mol}(\theta,\phi,R)^\ast
 D_e^{mol}(R) \cos\theta  \times  \\
& \psi_{r_m,\nu_m}^{mol}(\theta,\phi,R)R^2 \sin\theta dRd\theta d \phi,
\end{split}
\end{equation}
where $\psi_{r_n,\nu_n}$ and $\psi_{r_m,\nu_m}$ are the remaining ro-vibrational wave functions obtained after the integration over electronic
coordinates in Eq.~\eqref{eq:DefElectronicdipolemoment}. Now, we assume that the ro-vibrational wave function may be written as a product,
$\Psi_{r_n,\nu_n}(R,\theta,\phi)=\Phi_{r_n}(\theta,\phi)f_{\nu_n}(R)$. Then

\begin{equation} \label{eq:DipoleZaxisSquared}
|\vect{D}_Z^{lab}|^2= \Big|L_{J_n, M_n}^{J_m, M_m}\Big|^2 \times \Big|\int f_{\nu_n}(R) D_e^{mol}(R) f_{\nu_m}(R)R^2dR\Big|^2,
\end{equation}
with
\begin{equation} \label{eq:DefL}
L_{J_n, M_n}^{J_m, M_m}=\int {\Phi_{r_n}^{lab}}^\ast (\theta,\phi) \cos \theta \Phi_{r_m}^{lab}(\theta,\phi) \sin\theta d\theta d\phi.
\end{equation}
Defining
\begin{equation} \label{eq:DefHonlLondon}
S_{J_m,J_n}=3 \sum_{M_n,M_m} \Big|L_{J_n, M_n}^{J_m, M_m}\Big|^2
\end{equation}
known as the H\"{o}nl-London factors \cite{Honllondon,Kovacs,herzberg}. We combine the above results with Eq.~\eqref{eq:DSumSubstates} to find
the total transition dipole moment entering Eq.~\eqref{eq:AppQMEinstein}
\begin{equation} \label{eq:App_DSummedOverM}
|\boldsymbol{\mathcal{D}}_{m,n}|^2=S_{J_m,J_n}\Big|\int f_{\nu_n}(R) D_e(R) f_{\nu_m}(R)R^2dR\Big|^2.
\end{equation}

\subsection{Hunds case (a)}
We use the Hunds case (a) eigenfunctions in the lab frame from \cite{AmJPhys_angularMomentumStatesofDiatomic_OnHonlLondoncalculation} (cf.
Eq.~\eqref{eq:caseaeigenfuncts})
\begin{equation}
\begin{gathered}
 \Braket{\{\vect{r}_i\}\vect{R}|nJM_J\Omega S \Sigma}=\\ \sqrt{\frac{2J+1}{8\pi^2}}\Braket{\{\vect{r}_i^\prime\},
R|n}\Ket{S\Sigma}\mathcal{D}_{M_J\Omega}^{J^\ast}(\alpha \beta \gamma),
\end{gathered}
\end{equation}
and write the \emph{l}th component of the \emph{k}th moment transition operator in the laboratory frame, $T_l^k$, as a similar rotation of the
operator working in the molecular rest frame
\begin{equation} \label{eq:dipolemomentoperatorrotated}
    T_l^k(\{\vect{r}_i\} \vect{R})=\sum_{\Lambda=-k}^k T_\Lambda^k (\{\vect{r}_i^\prime\},R)\mathcal{D}_{l\Lambda}^{k^\ast}(\alpha \beta \gamma).
\end{equation}
Combining the above equations and performing the integral over Euler angles, while writing the Wigner rotation functions as an expansion over
Clebsch-Gordan coefficients \cite{sakuraiModernQM}, one finds the dipole moment transition matrix elements $(k=1)$
\begin{equation}
\begin{gathered}
    \Bra{n^\prime J^\prime M_J^\prime}T_l^1(\{\vect{r}_i\},\vect{R})\Ket{n^{\prime \prime} J^{\prime \prime} M_J^{\prime\prime}}= \\
    \sqrt{\frac{2J^{\prime\prime}+1}{2J^\prime+1}} \sum_{\Lambda=-1}^1\Bra{n^\prime \nu^\prime}T_\Lambda^1\Ket{n^{\prime\prime}
    \nu^{\prime\prime}}\\ \Braket{J^{\prime\prime} M_J^{\prime\prime} 1l|J^\prime M_J^\prime}\Braket{J^{\prime\prime} \Omega^{\prime\prime} k
    \Lambda|J^\prime \Omega^\prime}.
\end{gathered}
\end{equation}
Summing over the projections of $\vect{J}$ and emission directions one finds the line strength
\begin{equation}
\begin{split}
 &\sum_{M_J^\prime,M_J^{\prime\prime}}|\Bra{n^\prime J^\prime M_J^\prime}T_l^1(\{\vect{r}_i\} \vect{R})\Ket{n^{\prime \prime} J^{\prime \prime}
    M_J^{\prime\prime}}|^2=\\
    &(2J^{\prime\prime}+1) |\Bra{n^\prime \nu^\prime}T_\Lambda^1\Ket{n^{\prime\prime}
    \nu^{\prime\prime}}|^2 \times \\
    & |\Braket{J^{\prime\prime}\Omega^{\prime\prime}1 (\Omega^\prime - \Omega^{\prime\prime})|J^\prime \Omega^\prime}|^2
    \delta(S^\prime,S^{\prime\prime})\delta(\Sigma^\prime,\Sigma^{\prime\prime}).
    \end{split}
\end{equation}
Finally we find the H\"{o}nl-London factors in Hunds case (a)
\begin{equation}
\begin{gathered}
S(J^\prime,J^{\prime \prime})=(2J^{\prime \prime}+1) \times \\
|\Braket{J^{\prime\prime}\Omega^{\prime\prime}1 (\Omega^\prime- \Omega^{\prime\prime})|J^\prime \Omega^\prime}|^2
\delta_{S^\prime,S^{\prime\prime}}\delta_{\Sigma^\prime,\Sigma^{\prime\prime}}.
\end{gathered}
\end{equation}

\subsection{Hunds case (b)}
We gave the Hunds case (b) eigenfunctions in the laboratory frame in Eq.~\eqref{eq:caseBeigenfuncts}
\begin{equation}
\begin{gathered}
\Braket{\{\vect{r}_i\}\vect{R},nJM_J N M_N S M_S}=\sqrt{\frac{2N+1}{8\pi^2}} \times \\ \sum_{M_S=-S}^S \sum_{M_N=-N}^N
\Braket{\{\vect{r}_i^\prime\},R|n}\Braket{N M_N S M_S| J M_J} \times \\
\Ket{S M_S} \mathcal{D}_{M_N\Lambda}^{N^\ast}(\alpha \beta \gamma).
\end{gathered}
\end{equation}
The rotated dipole moment operator was given in a general form in Eq.~\eqref{eq:dipolemomentoperatorrotated}. We then use the identities
\cite{BrinkAndSatchlerAngularMomentumBook}
\begin{equation} \label{eq:WignerMatrixIdentities}
\begin{split}
\mathcal{D}_{l \lambda}^{k}\mathcal{D}_{m \mu}^{n}=&\sum_{N' M' \mu'} \Braket{n m k l|N' M'} \times \\ & \Braket{n \mu k \lambda|N' \mu'}
\mathcal{D}_{M \mu'}^{N'}
\end{split}
\end{equation}
and
\begin{equation}
\int \mathcal{D}_{l m}^{k}\mathcal{D}_{\lambda \mu}^{\kappa} d\Omega =\frac{8\pi^2}{2k+1}\delta_{l,\lambda}\delta_{m, \mu}\delta_{k,\kappa},
\end{equation}
where $\int d\Omega=\int_0^{2\pi} d\alpha \int_0^{2\pi} d\gamma \int_0^\pi d\beta sin\beta$. One thereby finds the expression for the dipole
matrix element
\begin{widetext}
\begin{equation}
\begin{gathered}
 \Bra{n' J' M_J'}T_l^1(\{\vect{r}_i'\},\vect{R}) \Ket{n''J''M_J''}= \sqrt{\frac{2N''+1}{2N'+1}} \Bra{n' \nu'}T_{\Lambda'-\Lambda''}^1 \Ket{n'' \nu''}
\Braket{N''\Lambda''1(\Lambda'-\Lambda'')|N'\Lambda'}\times \\  \sum_{\underset{ M_S' , M_S'}{ M_N', M_N''}} \Braket{N''M_N'' 1 l|N' M_N'}
\Braket{N' M_N' S' M_S'|J'M_J'} \Braket{N''M_N''S''M_S''|J''M_J''}\delta_{S'S''}\delta_{M_S' M_S''}.
\end{gathered}
\end{equation}
\end{widetext}


This is summed over the projections of $\vect{J}$ and squared to find the dipole transition probability. The task is simplified by rewriting the
products of Clebsch-Gordan coefficients in terms of Wigner 6j symbols \cite{Sobelman}. After some algebra one then finds
\begin{equation}
\begin{gathered}
|\Bra{n'J'N'}T_l^1 \Ket{n''J''N''}|^2=\frac{1}{3} (2N''+1)\times \\
(2J'+1) (2J''+1) \Bra{n'\nu'}T_{\Lambda'-\Lambda''}^1\Ket{n''\nu''}^2 \times \\
\Braket{N'' \Lambda'' 1 (\Lambda'-\lambda'')|N' \Lambda'}^2  \begin{Bmatrix} S & N'' & J''\\ 1 & J' & N' \end{Bmatrix}  ^2.
\end{gathered}
\end{equation}
Summing over the emission directions cancels the factor of $\frac{1}{3}$, leaving the expression for the Hönl-London factor in Hunds case (b)
\begin{equation}
\begin{gathered}
S(J',J'')=(2N''+1)(2J'+1)\times \\ (2J''+1) \Braket{N'' \Lambda'' 1 (\Lambda'-\Lambda'')|N' \Lambda'}^2 \times \\ \begin{Bmatrix} S & N'' &
J''\\ 1 & J' & N'
\end{Bmatrix}^2 \delta_{S'S''}.
\end{gathered}
\end{equation}


\end{document}